\documentclass[12pt]{article}

\usepackage{indentfirst}
\usepackage{amsmath}
\usepackage{hyperref}
\usepackage{amssymb}
\usepackage{amsfonts}
\usepackage{amscd}
\usepackage{amsbsy}
\usepackage{amsthm}
\usepackage{latexsym}
\usepackage{graphicx,color} 
\usepackage{booktabs}

\def\nn{\nonumber} 
\def\n{\label}

\renewcommand{\vec}[1]{{\bf #1}}

\def\beq{\begin{eqnarray}}
\def\eeq{\end{eqnarray}}
\def\ln{\,\mbox{ln}\,}

\def\al{\alpha}
\def\be{\beta}

\def\ga{\gamma}
\def\de{\delta}
\def\vp{\varepsilon}

\def\ze{\zeta}

\def\na{\nabla}
\def\pa{\partial}

\def\si{\sigma}

\def\Ga{\Gamma}

\def\La{\Lambda}

\def\Om{\Omega}

\renewcommand{\vec}[1]{{\bf #1}}

\def\CC{cosmological constant}

\begin{document}

\begin{center}

{\large\bf Primordial Universe with the running cosmological constant}
\vskip 6mm

\textbf{Jhonny A. Agudelo Ruiz}$^{a,d}$
\footnote{E-mail address: \ jaar@cosmo-ufes.org},
\ \
\textbf{Tib\'erio de Paula Netto}$^{c}$
\footnote{E-mail address: \ tiberio@sustech.edu.cn},
\ \
\\
\textbf{J\'ulio C. Fabris}$^{a,b}$
\footnote{E-mail address: \ julio.fabris@cosmo-ufes.org},
\ \
\textbf{Ilya L. Shapiro}$^{d}$ 
\footnote{On leave from Tomsk State Pedagogical University.
E-mail address: \ shapiro@fisica.ufjf.br}
\vskip 3mm

$^{a}$
\textsl{N\'ucleo Cosmo-ufes \& PPGCosmo, Departamento de
F\'isica, Universidade Federal do Esp\'irito Santo, Vit\'oria,
29075-910, ES, Brazil}

$^{b}$
\textsl{National Research Nuclear University MEPhI,
Kashirskoe sh. 31
\\
Moscow, 115409, Russia}

$^{c}$
\textsl{Departament of Physics, Southern University of Science
and Technology
\\
Shenzhen, 518055, China}

$^{d}$
\textsl{Departamento de Física, ICE,
Universidade Federal  de Juiz  de  Fora,
\\
Juiz de Fora, 36036-100, MG, Brazil}
\vskip 2mm



\end{center}
\vskip 6mm

\centerline{{\bf Abstract}}

\begin{quotation}
\small
\noindent
Theoretically, the running of the cosmological constant in the IR
region is not ruled out. On the other hand, from the QFT viewpoint,
the energy released due to the variation of the cosmological constant
in the late Universe cannot go to the matter sector. For this reason,
the phenomenological bounds on such a running are not sufficiently
restrictive. The situation can be different in the early Universe
when the gravitational field was sufficiently strong to provide an
efficient creation of particles from the vacuum. We develop a
framework for systematically exploring this possibility. It is
supposed that the running occurs in the epoch when the Dark Matter
already decoupled and is expanding adiabatically, while the usual
matter should be regarded
approximately massless and can be abundantly created from vacuum
due to the decay of vacuum energy. By using the handy model of
Reduced Relativistic Gas for describing the warm Dark Matter,
we consider the dynamics of both cosmic background and linear
perturbations and evaluate the impact of the vacuum decay on the
matter power spectrum and to the first CMB peak. Additionally,
using the combined SNIa+BAO data, we find the best-fit values
for the free parameters of the model.
\vskip 2mm

\noindent
{\sl Keywords:} \ Early Universe, running cosmological constant, first CMB peak
\vskip 2mm

\end{quotation}



\section{Introduction}

The improving quality of the data of observational cosmology leads
to better estimates of the equation of state of the Dark Energy,
which is driving the accelerated expansion of the Universe. The
current data are consistent with the value of $w=-1$, which means
the cosmological constant. From the quantum field theory point of
view, the cosmological constant is a necessary element of a
consistent semiclassical theory
\cite{UtDW, nelspan82, book, PoImpo} and hence
it should not be taken as a surprise that it is non-zero.

On the other hand, the ultimate word about the origin of the Dark
Energy belongs to observations. It can not be ruled out that at some
moment the analysis of the data proves that the density of the Dark
Energy changes with time. Does this mean that there is another
component of the Dark Energy, besides the cosmological constant?
Before answering this question, one has to understand whether the
cosmological constant can be not exactly a constant. It is a standard
assumption that the observable density of the vacuum energy is a sum
of the vacuum counterpart and the contribution generated by a
symmetry breaking, e.g. at the electroweak and QCD scales. In
principle, both vacuum and induced parts can be variable due to
quantum effects.

The variation of cosmological ``constant'' term, because of the
quantum effects, can be explored employing the renormalization
group running of this parameter \cite{CC-nova, DCCrun}. The simplest
version of such a running can be described in the framework of a
minimal subtraction scheme in curved space \cite{nelspan82, Buch84}
(see also \cite{book}), but this kind of running leads to the inconsistent
cosmological model \cite{CC-nova}. The standard interpretation
is that the ``correct'' running at low energies (in the IR) should
take into account the decoupling of the massive fields. Such
decoupling cannot be verified for the \CC\ case \cite{apco},
but the non-running can be proved neither \cite{DCCrun}. Thus, the
situation is such that one can explore the running \CC\ only in the
phenomenological setting. However, it is important to have this
setting well-defined. And in this respect, the main point is what
happens with the energy when the \CC\ varies according to the
evolution of the Universe and the corresponding change of the
energy scale.

It is well-known that the quantum or semiclassical corrections to
the action of gravity are typically non-local and rather complicated
(see e.g. \cite{apco}). However, one can identify the terms
responsible for the running of the cosmological constant using the
global scaling arguments  \cite{PoImpo}. Starting from this point,
one can meet two distinct possibilities to implement the \CC\
running in cosmology. The first one assumes the energy exchange
between vacuum and matter sectors. The cosmological model
that emerges from this assumption has essential technical
advantages. In particular, the evolution of the cosmological
background can be easily described using elementary functions
\cite{CC-fit} and the analysis of perturbations is also relatively
simple \cite{CCwave}. For this reason, this model became popular
(see, e.g, the review \cite{Sola-CC-review} and the recent
publication \cite{BMS1,BMS2}), regardless of the existing
conceptual difficulties, that will be described below. The second
model is much more consistent for the low-energy regime,
it is based on the conservation law not involving the matter sector,
and assumes a mixture between the \CC\ term and the Einstein-Hilbert
action, that means a running of the Newton constant $G$. This model
is more complicated technically, and also the phenomenological
restrictions on the unique free parameter $\nu$ are very weak,
at least from the analysis of structure formation
\cite{CCG}\footnote{In compensation, running $G$ has
interesting astrophysical applications (see e.g.
\cite{RotCurves, Davi}).}. In what follows, we shall concentrate
on the models of running \CC\ of the first kind and explore the
physical conditions where this model makes sense.

Two clarifying observations considering the definition of our model,
are in order at this point. First of all, due to the Planck suppression,
the fourth- and higher-derivative terms in the classical action and loop
corrections are irrelevant even at the relatively high energy scale,
such as the one we deal with in this paper. To understand this, let us
quote the Starobinsky model of inflation \cite{star,star83}. This
model  is mostly based on the higher derivative $R^2$-addition to the
Einstein-Hilbert action. In the presence of anomaly-induced terms,
there is a solution with a constant Hubble parameter $H$ \cite{star}.
However, in the course of inflation $H$ decreases (approximately
linearly with time) and its magnitude at the end of the inflationary
epoch is supposed to be about $10^{13}\,GeV$. This provides a
sufficient difference between the effect of the intensive running of
the cosmological constant which we shall explore and the effect of
the $R^2$-term. Indeed, the running $\La$ in the presence of the
$R^2$-term may be relevant in earlier phases of inflation, but
this is another issue to study and we leave it to the future work.

Assuming the simultaneous energy exchange between the
cosmological constant density and Einstein-Hilbert sector
and between the cosmological constant density and matter
sector leads to an ambiguity in the cosmological model. Due
to this feature, the models with double energy exchange were
never elaborated, regardless on an extensive literature on the
running cosmological models (see e.g. \cite{Sola-CC-review}).
Besides the mentioned ambiguity, from the practical side there is
no much sense in considering such a double energy exchange,
because the effect of the running of $G$ is known to be much
weaker than the one of the energy exchange with matter
\cite{CCG}.

The main problem with the model based on the vacuum-matter
energy exchange is that during most of the history of the Universe
the typical energies of the gravitational degrees of freedom are
very small compared to the masses of all known particles
\cite{OphPel}. For instance, the value of the Hubble parameter
today is about  $H_0 \propto 10^{-42}\,$GeV, while the lightest
neutrino is supposed to have the mass about thirty orders of
magnitude greater. Thus, there is only a possibility to create
photons and this is not phenomenologically interesting,
since the energy density of such photons would be about $T^4$,
with the temperature $T \approx H$. Such an energy density is
of course much smaller than the energy density of CMB, which
is yet about four orders of magnitude smaller compared to the
present-day critical density, or to the \CC\ density. This argument
represents a serious obstacle to using this model for a {\it late}
cosmology.

Let us note that the described restrictions do not apply to the
{\it early} \ Universe, e.g., to the epoch after inflation, where the
value of the Hubble parameter is decreasing from about
$10^{13}\,$GeV to the values that are comparable to the
energy scale of the Minimal Standard Model of elementary particle
physics. This is a reheating period, where the creation of particles
is very intensive, and there is nothing wrong with assuming that this
happens because of the decay of the cosmological constant into the
matter.  In the next section, we shall explore the model \cite{CC-fit}
in the high energy domain. The description of quantum effects is
based on the running of \CC\ described in this paper. At the same
time, the models of early Universe require special care about the
description of matter. The matter contents of the Universe consist
mainly of the usual matter particles and DM. We assume that the
DM consists of the GUT
remnants and hence has masses that are much greater than the value
of $H$. Thus, the DM can be regarded to decouple, in the sense
that DM particles are not created from the vacuum. Thus, an
appropriate description of DM is an ideal gas of massive particles
adiabatically expanding and becoming less relativistic with
time. To describe such a gas, we shall use the simple and
convenient Reduced Relativistic Gas (RRG) model, which was originally
developed by Sakharov in the classical paper \cite{Sakharov1966},
and recently reinvented in \cite{Sobrera2005, Sobrera2009}.

The paper is organized as follows. In the next section, we formulate
the framework and the model, including the Einstein-Hilbert action
with the running \CC\ and non-running Newton constant, DM
described by RRG, decoupled from everything except the standard
gravitational interaction, and the usual matter, that has the
equation of state of radiation and is exchanging energy with the varying
\CC\ sector. In Sec.~\ref{s3} we describe the perturbations in this
model, and derive the observable consequences of the running
\CC\ in Sec.~\ref{s4}. Finally, in  Sec.~\ref{s5} we draw our
conclusions and discuss the perspectives for subsequent work.

\section{Background solution}
\label{s2}

We consider a cosmological model with the possibility of particle
creation in the primordial Universe due to the quantum effects of
vacuum. More precisely, we study the vacuum energy decay as a
result of the renormalization group (RG) equation for the density
of the \CC\ term.

In Refs.~\cite{CC-nova,Babic2002,CC-fit,CC-Gruni} 
it has been shown, from the general arguments based on covariance 
and dimensions, that the form of these quantum corrections can be 
defined up to a single free parameter $\nu$,
\beq
\n{eq-La}
\rho_\La \,=\, \rho^0_\La \,+\,
\frac{3\nu}{8\pi G}\, \big( H^2 - H_0^2),
\label{CCrun}
\eeq
where the subscript $0$ means that the quantity is taken at some
reference redshift parameter when time is $t_0$ and the conformal
factor $a_0$. Regardless the formula (\ref{CCrun}) is non-covariant,
the main argument of \cite{CC-nova,Babic2002} (see also
\cite{PoImpo}) was based on covariance, and
looks as follows. The effective action terms which can be classified
as quantum contributions to the \CC\ are certainly non-local, but
they are also covariant. Making an
expansion in the powers of metric derivatives (on flat or even
de Sitter background), we arrive at the local expressions and all
the terms in these expansions are of the {\it even} powers in metric
derivatives. The reason is that with the odd powers it is
algebraically impossible to provide a scalar term in the effective
action, regardless of the level of complexity of the non-local
action. On the cosmological background, the absence of odd powers
of derivatives means that the first terms of the expansion include
$H^2$ and ${\dot H}$. Now, the possible
${\cal O}({\dot H})$-actions are surface terms or they reduce to
${\cal O}(H^2)$ when
substituted into the action\footnote{Indeed, this argument has no
absolute power because this and other terms can emerge on the
way from the action to equations of motion. On the other hand,
phenomenologically ${\cal O}({\dot H})$-term is also not very
relevant \cite{Sola-CC-review}.}.

As an additional discussion of the motivations of our work, let us
mention that the same formula for the running (\ref{CCrun}) follows
from the hypothesis of quadratic decoupling for the cosmological
constant \cite{Babic2002,CC-fit}. In the case of cosmological or
Newton constants, this is just a well-motivated hypothesis, while for
the fourth-derivative terms it can be proved by direct quantum
calculations \cite{apco,fervi,Omar-FF4D}. It is important to note
that the identification of the scaling (renormalization group) parameter
$\mu$ with the Hubble parameter $H$ has been achieved not only on
the phenomenological background, but was also derived within the
especially developed scale-setting procedure \cite{Babic2005}.
A similar procedure has been later on successfully used in
astrophysical applications \cite{StefDom} to confirm the guess
\cite{RotCurves} based on the qualitative quantum field theory
based arguments (see also recent work \cite{BWD} for a more
sophisticated astrophysical and cosmological considerations and
further references).

Let us especially stress that the  (\ref{CCrun}) is \ {\it not} based
on primitive dimensional considerations. Indeed, dimensional
arguments can be used in similar frameworks (see e.g.
\cite{Farina}), but it makes sense only when combined with sound
quantum field theory-based, or purely phenomenological, arguments.
As an example of the primitive use of dimension, let us mention
the recent papers \cite{SSP}\footnote{Citations and comment
included by the request of the anonymous referee.}. The mentioned
set of papers is based on the identification of the variable
cosmological constant in the form
\beq
\La(t)\,=\,\La_{bare} + \frac{\alpha^2}{t^2},
\label{at}
\eeq
with dimensionless $\al$,
which is called a running cosmological term. The term ``running''
and the notation $\La_{bare}$ imply some relation to quantum theory,
but in fact, the formula (\ref{at}) is based only on the coincidence of
the dimension of $\La$ and the inverse dimension of the square of
the cosmological time coordinate $t$. The phenomenological effect
of the variable cosmological constants formulas (\ref{CCrun})
and (\ref{at}) should be expected to be similar, because in the
matter-dominated or radiation dominated models we have
$H\sim t^{-1}$. Due to the presence of the constant components
($\La_{bare}$ and $\rho^0_\La$, correspondingly) the formulas are
not equivalent, but for appropriately chosen $\nu$ and $\al$ the
phenomenological effects of the ``running'' may be similar. The
main difference between the two formulas is conceptual. In fact,
(\ref{at}) implies that one of the fundamental parameters is a
given function of the time coordinate.  In other words, the
cosmological constant is not an independent quantity or field (such
as, e.g., quintessence), but a subject of an extra force acting on all
the Universe from some kind of external source. It is important that
this source is external to the whole our Universe, as otherwise one
can not guarantee the given time-dependence being independent on
the matter contents of the Universe and other parameters, defining
the dynamics of the conformal factor and perturbations.

It is clear that introducing an appropriately fine-tuned external
force acting on all the Universe one can provide many desirable
features in the observables (see e.g. \cite{WolframElph} for
some examples of this sort\footnote{Citation recommended by
anonymous referee.}). However, taking any result of this
kind seriously would contradict the existing in Natural Sciences
tradition of trying to explain observables from some fundamental
principles (in the extent we are capable to formulate them, of
course) and not using a specially fine-tuning external source to
explain one or another set of observables. In this respect, the
formula (\ref{CCrun}) is completely different from the formula
(\ref{at}). In the first example the dynamics of the cosmological
constant density is defined by the solution of the equations of
motion, and not defined by hand from the primitive dimensional
arguments, as it is the case in the last example.

Another relevant observation concerns the equation of state of the
running cosmological constant density. In the model formulated
in \cite{CC-fit}, and in what follows, it is assumed that this
equation remains the same as for the non-running CC. The main
reason for this is that the assumed quantum corrections which
contribute to the running have the same global scaling as the
constant term, or as the $R\Box^{-2}R$ and other similar terms,
as it was discussed in \cite{apco} (see also \cite{DCCrun}).

At the sufficiently low energy scale, the next terms in the expansion
in derivatives, such as ${\mathcal O}(H^4)$ or
${\mathcal O}({\dot H}^2)$, are negligible compared to the quadratic
term in  (\ref{CCrun}). The epoch which we intend to deal with in the
present paper is the end of inflation and the transition to the
radiation-dominated Universe. The standard estimate for the initial
value of $H$ in this epoch is $H \propto 10^{11}-10^{12}\,GeV$. Then
the four-derivative terms are suppressed by factor of
$10^{-14}-10^{-10}$ compared to the quadratic in derivative terms.
Thus, it is sufficient to explore the cosmological models based on the
relation (\ref{CCrun}).

In what follows primes indicate derivatives with respect
to the redshift parameter
\beq
1 + z = \frac{a_0}{a}.
\label{z}
\eeq
In the present paper, we use the normalization with the scale factor at
present $a_0=1$. The sign of $\nu$ indicates whether bosons or
fermions dominate in the running \cite{CC-fit}.

The matter contents of the Universe include usual matter, DM and
radiation, according to the current estimate \cite{Planck2015I}. Here,
the DM component is described as a reduced relativistic gas (RRG) of
massive particles, which take into account in a simple and useful way
the warmness of the fluid. The RRG is a reliable approximation when
the interaction between the particles is irrelevant  \cite{Sobrera2005}.
As the main simplification compared to the J$\ddot{\rm u}$ttner
model \cite{Juttner}, RRG assumes that the particles composing
the have equal speeds, $v = c \be$. As we have explained above, for
the usual matter we assume an ultrarelativistic equation of state
with $P_{b}\approx \frac13\rho_{b}$.

An elementary consideration \cite{Sobrera2005} (see also
\cite{Simpl,Leo} for alternative derivations) shows that the
equation of state of such a gas is
\beq
\n{EoSm}
P_{dm}
\,=\, \frac{\rho_{dm}}{3}
\bigg[1 -  \Big(\frac{m c^2}{\vp} \Big)^2\, \bigg]
\,=\, \frac{\rho_{dm}}{3}
\Big(1 - \frac{\rho_d^2}{\,\rho_{dm}^2}\,\Big),
\eeq
where $\vp = \frac{mc^2}{\sqrt{1-\be^2}}\,$ is the kinetic energy
of the individual particle, $\rho_{dm}=n\vp$ and $P_{dm}$ are energy
density of the gas, while $\rho_d=nmc^2$ is the density of the rest
energy. Consequently, the scaling rule for this quantity is
\beq
\rho_d(z) = \rho_{d}^0 (1+z)^3.
\label{rhod}
\eeq
Here we consider an early post-inflationary Universe, where the
DM has already decoupled from the other matter components
and satisfies a proper continuity equation
\beq
\n{eq-rho2}
\rho_{dm}' = \frac{(4-r)}{1+z} \, \rho_{dm},
\eeq
where we introduced the useful function
\beq
r \,=\, r(z) \,=\, \frac{\rho^2_d(z)}{\rho^2_{dm}(z)}.
\eeq
In the early Universe, one can restrict the consideration by
the spatially flat FLRW metric. The solution for Eq.~(\ref{eq-rho2})
can be easily found for a single adiabatically expanding fluid
\cite{Sobrera2009}. Then the scaling law for the relative energy
density (relative to the critical density today\footnote{Here and from now on we use the notation $\Omega_i (z)=\rho_i (z)/\rho^{0}_c\,$, where $\rho^0_c \,=\, 3 H_0^2/8 \pi G$.}) for the relativistic gas
representing the DM, is given by the expression
\beq
\n{Om}
\Om_{dm}(z)
\, = \,
\frac{\Om_{dm}^0(1+z)^3}{\sqrt{1+b^2}} \, \sqrt{1+b^2(1+z)^2},
\qquad
\mbox{where}
\qquad
b = \frac{\be}{\sqrt{1-\be^2}}
\eeq
and $\Om^0_{dm}$ is the DM density in the present-day Universe.
The parameter $b$ measures the warmness of the matter (DM in our
case).  For a low warmness $\,\be \ll 1$, we have $\,b \sim \be $.
Thus, $\,b \approx  0$ means that the matter contents is ``cold''.
The RRG model provides an interpolation between the radiation
($b \to \infty)$ and matter ($b=0$) dominated regimes
\cite{Sobrera2005}. The model can be used also to describe several
fluids in the thermal contact \cite{Zimdahl2014, Hipolito2017}.

According to our physical setting, the running \CC\ \cite{DCCrun}
is exchanging energy only with the usual matter and the last
has the approximate equation of state of radiation. Then the conservation
law has the form
\beq
\n{eq-rho1}
 \rho_r' - \frac{3(1+w)}{1+z} \, \rho_r = - \rho_\La',
\eeq
where we left $w$ to be the equation of state parameter for the sake of
generality. When starting to deal with the numerical estimates,
we shall set $w=1/3$. Finally, the Hubble parameter is given
by the Friedman equation
\beq
\label{eq-H}
H^2 (z) \,=\, \frac{8\pi G}{3} \big[
\rho_\La (z) + \rho_r (z) + \rho_{dm} (z)\big].
\eeq

The solution of the system (\ref{eq-La}), (\ref{eq-rho1})
and (\ref{eq-H})
can be performed following the pattern of \cite{Espana2004}, since
the technical complications related to the presence of DM are not
critical. In order to obtain $\Om_r(z)$ one has to consider the
derivative of Eq.~(\ref{eq-H}) and then use (\ref{eq-La}). After this,
we arrive at the equation
\beq
\n{rola1}
\rho_\La'  \,=\,  \frac{\nu}{1-\nu} \big(\rho'_r + \rho'_{dm}\big).
\eeq
Using (\ref{rola1}) in (\ref{eq-rho1}) to eliminate $\rho_\La$, after some
simple algebra we obtain the differential equation for $\rho_{r}(z)$,
\beq
\n{eq-rb}
\rho_r' - \frac{\ze}{1+z} \rho_r  \,=\,  - \nu \rho_{dm}',
\eeq
where
\beq
\ze  \,=\,  3(1+w)(1-\nu).
\label{zeta}
\eeq
Let us stress that the interaction between radiation (remember
it is all usual matter in this case) and DM, is not direct, but
occurs because of the running of the \CC\ term in Eq.~(\ref{CCrun}),
parameterized by $\nu$, and the Friedmann equation (\ref{eq-H}).
This implicit interaction occurs regardless of the DM satisfies
separate continuity equation (\ref{eq-rho2}).

Using Eq.~(\ref{Om}) the solution of (\ref{eq-rb}) can be found in the form
\beq
&&
\Om_r (z) \,=\, C_0 (1+z)^\ze
- \frac{\nu \Om_{dm}^0(1+z)^3}{\sqrt{1+b^2}}
\Big[ \sqrt{1+b^2 (1+z)^2}
+ \frac{\zeta}{3-\zeta}\,\, {}_{2} F_1 (\al, \be; \ga; Z) \Big],
\mbox{\qquad}
\label{eq14}
\eeq
with
\beq
C_0 = \Om_{r}^0
+ \frac{\nu  \Om_{dm}^0}{\sqrt{1+b^2}}
 \Big[ \sqrt{1+b^2} +  \frac{\zeta}{3-\zeta } \,\,
{}_ 2F_1 (\al, \be; \ga;-b^2) \Big].
\eeq
Here \ ${}_2 F_1(\al, \be; \ga; Z)$ is the hypergeometric function
defined as
\beq
{}_2 F_1(\al, \be; \ga; Z) = \sum_{k=0}^\infty \frac{(\al)_k (\be)_k}{(\ga)_k}
\,\frac{Z^k}{k!},
\eeq
where $(\al)_k$ is the Pochhammer symbol. In our case
\beq
\al = -\frac12,
\qquad
\be = \frac{3-\ze}{2},
\qquad
\ga = \frac{5-\ze}{2}
\qquad
\mbox{and}
\qquad
Z = -b^2(1+z)^2.
\eeq

Furthermore,
$\Om_\La(z)$ is directly obtained by integrating (\ref{rola1}),
\beq
\Om_\La(z) =  B_0 + \frac{\nu}{1-\nu} \, [\Om_r(z) + \Om_{dm}(z) ],
\eeq
where
\beq
B_0 \,=\,
\Om_{\La}^0 - \frac{\nu}{1-\nu} \, \big(\Om_{r}^0 + \Om_{dm}^0\big).
\eeq
Finally, the Hubble parameter can be found from the Friedmann equation,
\beq
H (z)  \,=\,  H_0 \sqrt{\Om_\La(z) + \Om_r(z) + \Om_{dm}(z)}.
\eeq

To illustrate the behavior of the model we can consider the total
effective equation of state. It can be obtained using the
second Friedman equation,
\beq
-2(1+z) H H' + 3H^2 \,=\, - 8 \pi G P_t \,\equiv\,
- 8 \pi G \, w_{eff} (z)  \rho_t,
\eeq
where
\beq
\rho_t (z) \equiv \rho_\La (z) + \rho_r (z) + \rho_{dm} (z).
\label{rhotot}
\eeq
Thus,
\beq
w_{eff}(z) =\frac{2H'}{3H} - 1.
\eeq

\begin{figure} [t]
\centerline{
\includegraphics[scale=0.55]{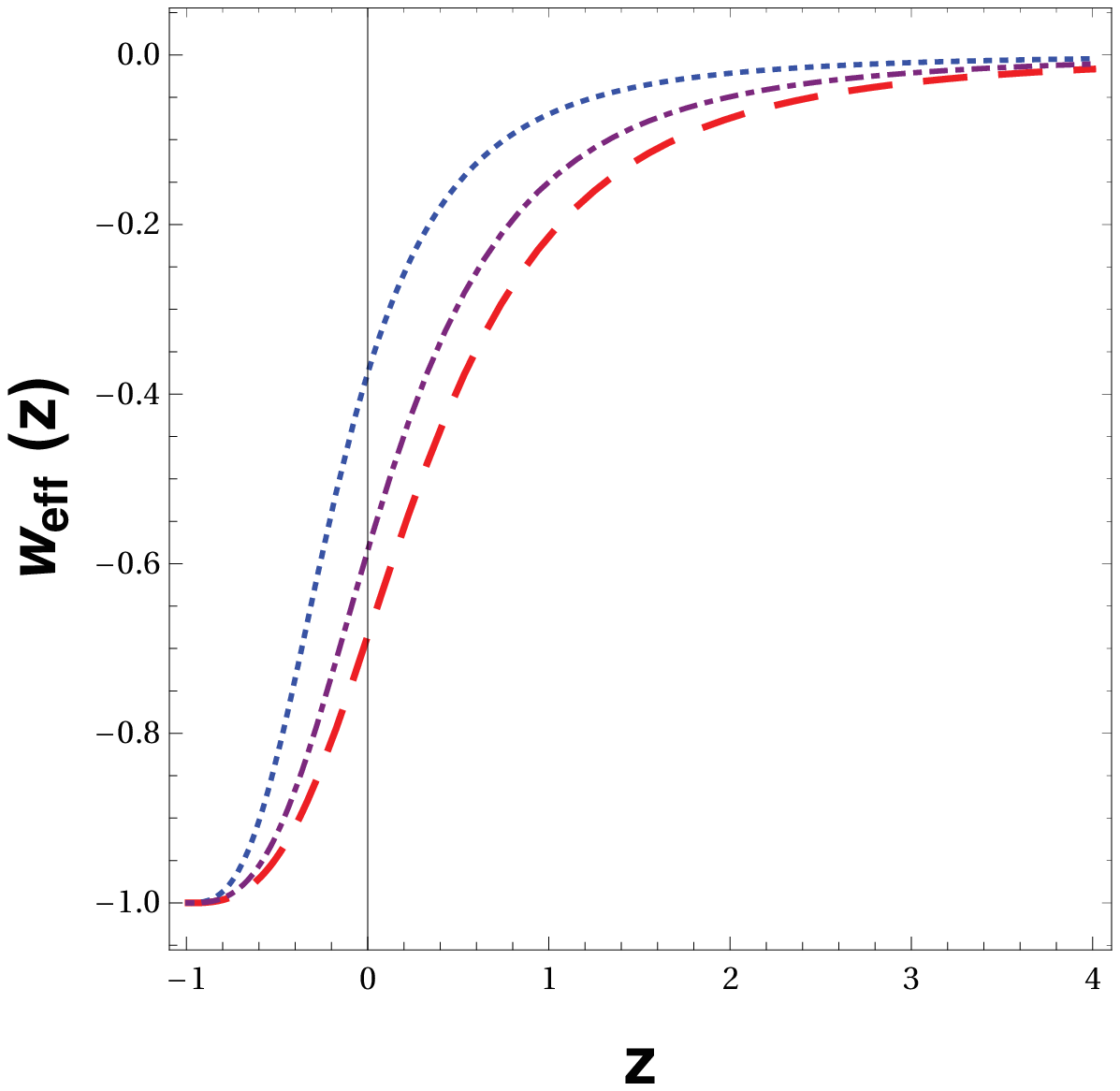}
\hskip 1.2cm
\includegraphics[scale=0.56]{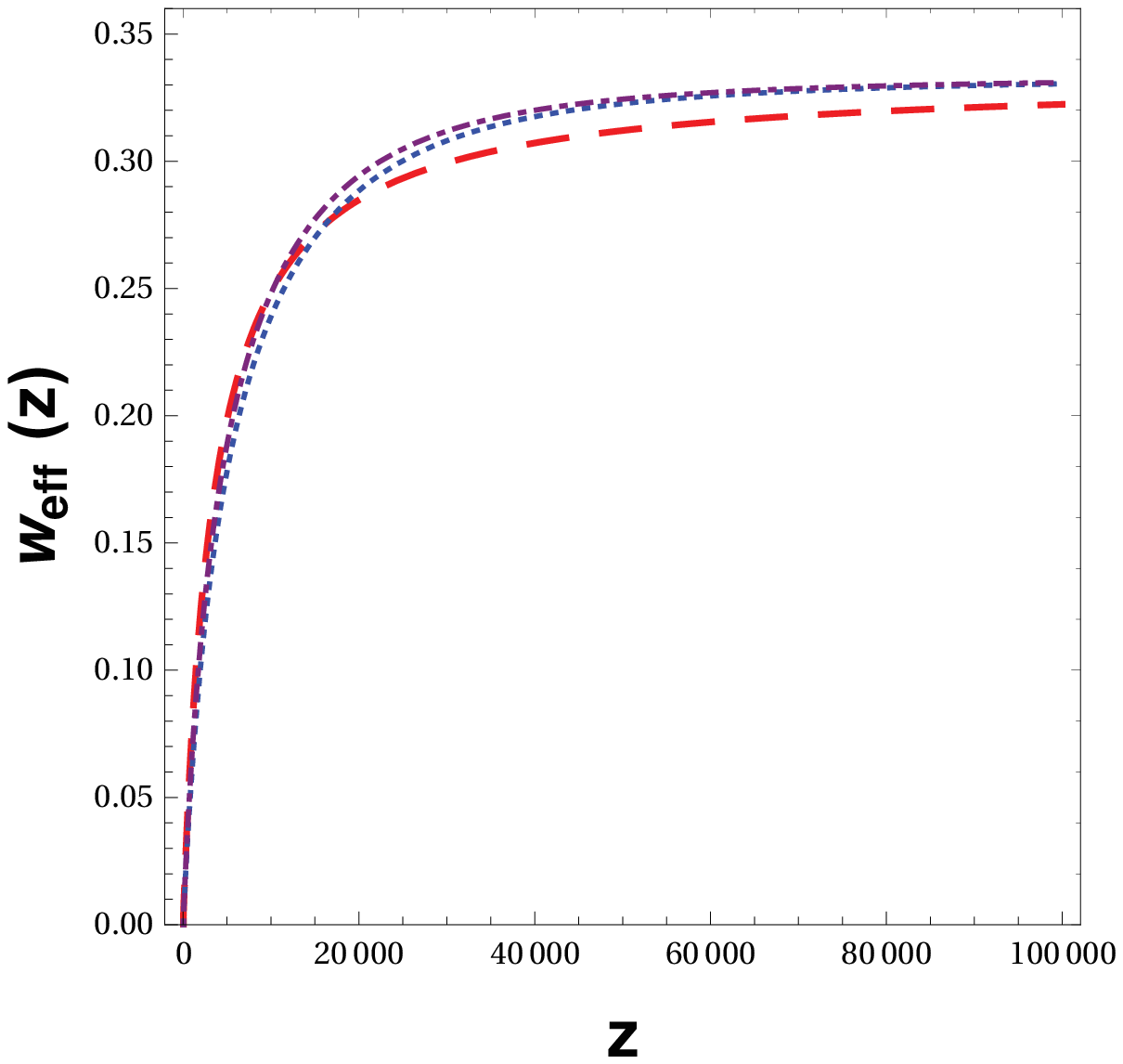}
}
\begin{quotation}
\caption{\small The dotted and dot-dashed lines represent our model for
$\chi_{l_1}^2$ and $\chi_{l_1}^2+\chi_{BAO}^2+\chi_{SNIa}^2$
best fits, respectively, based on the results from Sec.~\ref{s4}.
The dashed one represents the $\La$CMD model. On the left, we plot
$w_{eff}(z)$ for small values of $z$ and in the right plot there are
higher values of $z$. In the far future, $w_{eff}(z)$ approaches the
equation of state of constant $\La$. On the other hand, when
$z \to \infty$, the effective equation of state approaches radiation.}
\label{fig0}
\end{quotation}
\end{figure}

In Fig.~\ref{fig0}, we plot $w_{eff}(z)$ for the energy balance
obtained by the best fit of $\chi_{l_1}^2$ and
$\chi_{l_1}^2+\chi_{BAO}^2+\chi_{SNIa}^2$ (see
subsections \ref{CMBp} and \ref{Data} below).
As expected $w_{eff}(z) \to -1$ when $z \to -1$, while
$w_{eff}(z) \to \frac13$ for $z \to \infty$ \cite{Melchiorri2003}.
When compared with the $\La$CDM model with the same $\Om$'s and
CMB+BAO+SNIa combined data are used, our model fits better for
small $z$ and approaches faster to radiation dominated epoch when
$z$ increases.

\begin{figure} [t]
\centerline{
\includegraphics[scale=0.9]{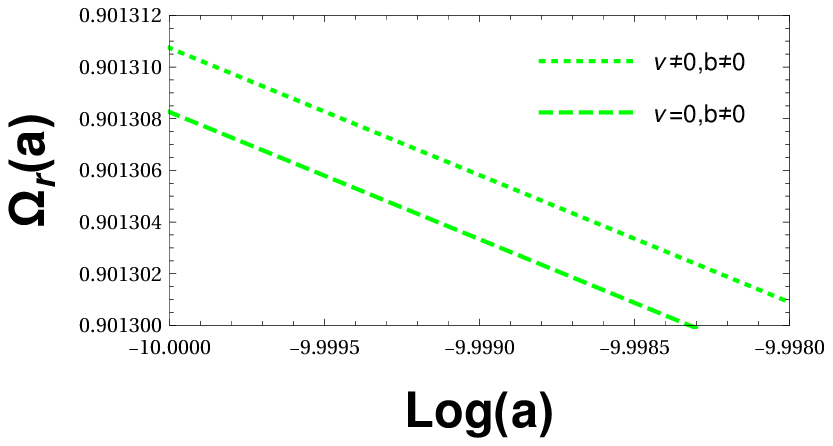}
\hskip 1.2cm
\includegraphics[scale=0.9]{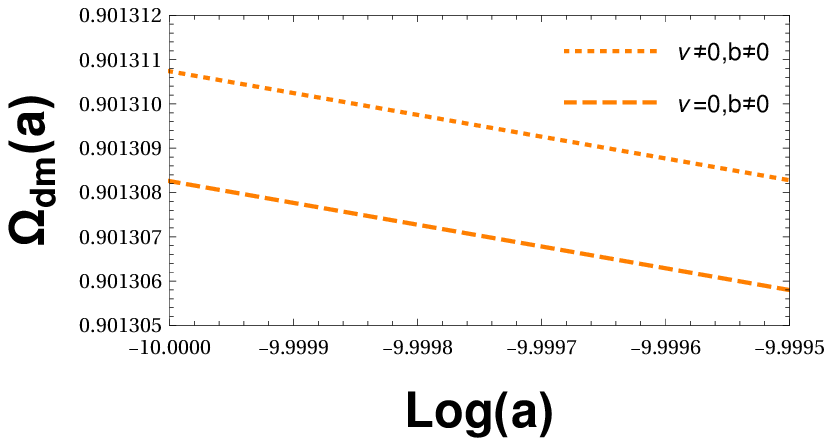}
}
\begin{quotation}
\caption{\small The relative densities for usual (baryonic)
matter
and DM, described by RRG. There is a very small difference
at the primordial epoch, compared to the $\La$CDM model.
This difference does not affect the BBN process.}
\label{fig0.1}
\end{quotation}
\end{figure}

It is worthwhile mentioning that the decay of the cosmological
term into radiation which includes relativistic (in the very early
Universe) matter, does not affect the nucleosynthesis
process, as it can be seen in Fig.~\ref{fig0.1}, where the abundance
of the relativistic species, including usual  (baryonic) matter is
compared with
the corresponding data of $\Lambda$CDM case represented by
$\nu = 0$. In this plot, we take again the best fit values given by
using CMB+BAO+SNIa combined data, this is,
$\nu =2.44\times 10^{-6}$ for the \CC\ running parameter
and $b = 1.89\times 10^{-6}$ for the DM warmness.

\section{Including perturbations}
\label{s3}

The cosmological perturbations in the model described above
can be analyzed following the approach developed in
Refs.~\cite{CCwave} and \cite{Sobrera2009}. This implies
 simultaneous perturbations of metric, energy density and the
 four-velocities in the co-moving coordinates,
\beq
g_{\mu\nu} \to g_{\mu\nu} + h_{\mu\nu},
\quad
\rho_i \to \rho_i + \de \rho_i,
\quad
U^\al \to U^\al + \de U^\al,
\quad
V^\al \to V^\al + \de V^\al\,,
\mbox{\quad}
\eeq
in the synchronous gauge $h_{0\mu} = 0$. Here $U^\al$ is the DM
velocity and $V^\al$ is the usual, or baryonic, matter (radiation, in
our case)
velocity. In the following calculations we use the constraint
$\de U^0 = \de V^0 = 0$.

The perturbation of the DM pressure should be derived from
the equation of state (\ref{EoSm}),
\beq
\de P_{dm}
\,=\, \frac{\de \rho_{dm}}{3}
\Big[ 1 - \Big(\frac{mc^2}{\vp} \Big)^2 \Big]
\,=\, \frac{\de \rho_{dm} (1-r)}{3},
\eeq
meaning that the perturbations satisfy the same equation of state as
the background quantities. Technically, this means that the variations
of the energy density $\de\rho_{dm}$ and the rest energy density
$\de \rho_d$ are always proportional. The reason for this restriction
is that in the framework of the RRG model one has to provide
kinetic energies of all particles to be equal and, therefore, we have
no right to change the ratio $\,\frac{mc^2}{\vp}$ \cite{Sobrera2009}.
The definitions of  the perturbations for other densities are
straightforward.

Let us introduce useful notations for the quantities related to
Eq.~(\ref{rhotot}),
\beq
f_1 (z) = \frac{\rho_r (z)}{\rho_t (z)},
\qquad
f_2 (z) = \frac{\rho_\La (z)}{\rho_t (z)},
\qquad
f_3 (z) = \frac{\rho_{dm} (z)}{\rho_t (z)},
\qquad
g(z) = \frac{2\nu}{3H(z)}.
\eeq
Thus, we arrive at the $00$-component of Einstein equations,
\beq
\n{eqh1}
h' - \frac{2h}{1+z}
\,=\, - \frac{2\nu}{(1+z)g} \big[
(1+3w)f_1 \de_r - 2 f_2 \de_\La + (2-r)f_3 \de_{dm}
\big],
\eeq
where $h = \pa_t (h_{ii}/a^2)$ and
\beq
\de_i = \frac{\de \rho_i}{\rho_i}
\eeq
are the corresponding density contrasts. Equations corresponding
to the time and spatial components of the perturbation for the
conservation law $\,\de (\na_\mu T^{\mu\nu}) = 0$, have the form
\beq
&&
\de_r'
\,+\, \Big[
\frac{f_1'}{f_1} - \frac{3(1+w)f_2}{1+z}
+ \frac{(1-r-3w)f_3}{1+z}
\Big] \de_r
\,-\,
\frac{1+w}{(1+z)H}
\Big(\frac{v}{f_1} - \frac{h}{2} \Big)
\nn
\\
&&
\,\,\quad
=\, -\,\frac{1}{f_1}(\de_\La f_2)'
\,-\, \frac{3(1+w)f_2}{1+z}
\Big[ 1 + \frac{(4-r)f_3}{3(1+w)f_1} \Big]
\de_\La,
\label{eqr1}
\\
\n{eqv1}
&&
v' + \frac{[3(1+w)f_1+(4-r)f_3-5]}{1+z} \, v
\,=\,
\frac{k^2(1+z)}{(1+w)H} \left(f_2 \de_\La - w f_1 \de_r \right),
\\
\n{1}
&&
\de'_{dm}
+ \Big\{
\,\frac{f_3'}{f_3}
+ \frac{3(1+w) f_1 + (r-4) (f_1 + f_2)}{1+z}
\Big\} \de_{dm}
\,+\, \frac{4-r}{3H(1+z)} \Big( \frac{h}{2} - \frac{u}{f_3} \Big)
\,=\, 0,
\mbox{\qquad}
\\
\n{2}
&&
u' + \Big[
\frac{3(1+w)f_1+ (4-r)f_3 -5}{1+z}
 - \frac{r'}{4-r}
\Big] u
+ \frac{k^2(1+z)f_3}{H} \Big( \frac{1-r}{4-r} \Big) \de_{dm}
\,=\, 0.
\mbox{\qquad\quad}
\eeq
Here we used the notations
$\,v = f_1 \na_i (\de V^i)\,$ and  $\,u = f_3 \na_i (\de U^i)\,$ for
divergences of the peculiar velocities and we rewrote all the
previous perturbation equations in the Fourier space, using
\beq
f(\vec{x},t)
\, = \,
\int \frac{d^3 k}{(2\pi)^3} \,  f(k,t) \,e^{i \vec{k} \cdot \vec{x} },
\quad
\mbox{with}
\quad
k = |\vec{k}|.
\eeq
Perturbing the formula (\ref{eq-La}), one finds
\beq
\n{deLa}
\de_\La = \frac{g}{f_2} \Big(\frac{v}{f_1}-\frac{h}{2} \Big).
\eeq
The last equation is not dynamical, representing a constraint that
can be replaced into other equations. Using (\ref{deLa}) in
(\ref{eqh1}), (\ref{eqr1}), and (\ref{eqv1}), we arrive at the
equations
\beq
\n{3}
&&
h' + \frac{2(\nu-1)}{1+z}\,h = \frac{2\nu}{1+z} \Big[
\frac{2v}{f_1}-(1+3w)\frac{f_1}{g} \de_r
- (2-r)\frac{f_3}{g} \de_{dm}
\Big],
\\
\n{4}
&&
\de_r' + \Big[
\frac{f_1'}{f_1} - \frac{3(1+w)f_2}{1+z}
+ \frac{(1-r-3w)f_3}{1+z}
\Big] \de_r
= \frac{1}{f_1}\Big(\frac{gh}{2} -\frac{gv}{f_1}  \Big)'
\mbox{\qquad}
\nonumber
\\
&&
\quad\,\,
+\, \frac{1+w}{1+z}
\Big[ 3g + \frac{(4-r)g f_3}{(1+w)f_1} - \frac{1}{H} \Big]
\Big(\frac{h}{2} -\frac{v}{f_1}  \Big),
\\
\n{5}
&&
v'
+ \Big\{\frac{[3(1+w)f_1+(4-r)f_3-5]}{1+z}
- \frac{k^2 g (1+z)}{(1+w)H f_1} \Big\} v
\nn
\\
&&
\quad\,\,
=\,
-\frac{k^2 g(1+z)}{2(1+w)H} \Big( h + \frac{2w f_1}{g}\,\de_r \Big).
\eeq
The complete system of perturbation equations includes
(\ref{1}), (\ref{2}), (\ref{3}), (\ref{4}) and (\ref{5}).

\section{Observational tests}
\label{s4}

The free parameters of the cosmological model for the early Universe
with running \CC\ and energy exchange between vacuum and matter can
be constrained from various observational tests. Thus, the general
framework of the model formulated above may have different
applications (one can see e.g. \cite{Fabris2012testing} for
the possibilities in a simpler model without \CC\ running). As a first
step, in the present section we consider 
the two tests, namely the position of the first acoustic peak of the
CMB power spectrum and the inclusion of SNIa+BAO combined data.

Let us note that the process of \CC\ decay into normal particles, as
discussed in the previous sections, is effective in the primordial
Universe, that is long before BBN. For this reason, we are allowed
to use the transfer function in the usual standard format. However,
this process leaves traces for the later epochs of the Universe
evolution, encoded in the values of parameters $\nu$ and $b$. In
this way, one can use the tests from the late phase of the Universe
for exploring the effect of running \CC \ in the earlier epoch.

The statistical analysis of the data starts with the $\chi^2$ functions,
constructed according to the general expression
\beq
\chi^2 (X^j)
\,=\, \sum_{i=1}^N \left[
\frac{\mu_i^{obs} - \mu_i^{th} (X^j)}{\si_i}\right]^2,
\eeq
where $N$ is the total number of observational data, $\mu^{th}_i$ are
the theoretical predictions depending on free parameters $X^j$, and
$\mu_i^{obs}$ represent the observational values with an error bar
given by $\si_i$. In our case the free parameters are $\,\nu$,
$\,\Om_{dm}^0\,$ and $\,b$. Let us remember that $\nu$ defines
the running of the vacuum energy, while $\,\Om_{dm}^0\,$
and $\,b\,$ describe the DM relative density and warmness. As usual,
$\,\Om^0_\La = 1 - \Om^0_{dm} - \Om^0_{b} - \Om^0_{r}$.
It is worth mentioning that here we are dealing with the late
Universe, hence usual matter (baryonic) and radiation contents are
separated.

The probability distribution function is constructed from $\chi^2$ as
\beq
P(X^j) = A e^{-\chi^2(X^j)/2},
\eeq
where $A$ is a normalization constant.

\subsection{The first CMB peak}
\label{CMBp}

\begin{figure}[t]
\centerline{
\includegraphics[scale=0.65]{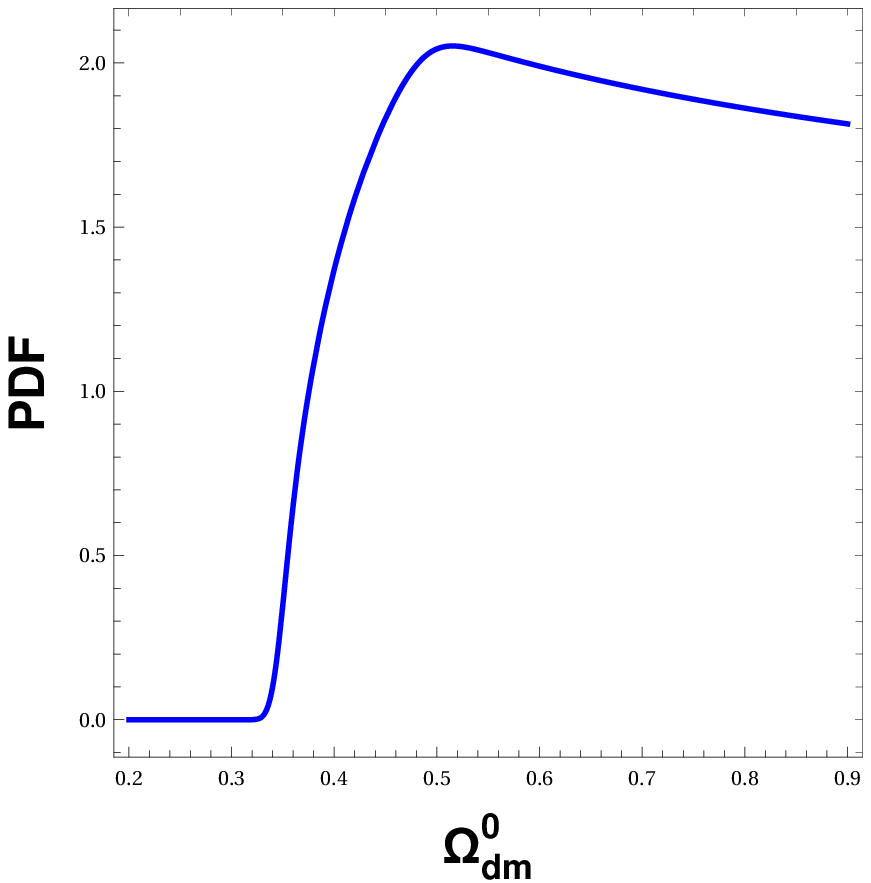}}
\centering{\includegraphics[scale=0.65]{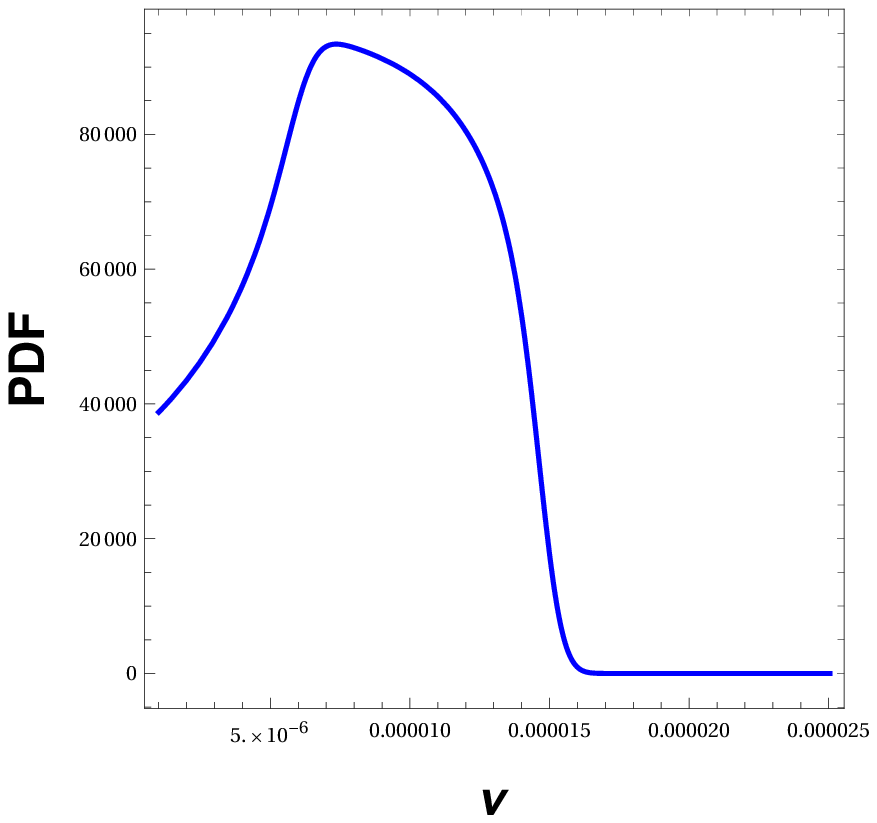}
\includegraphics[scale=0.65]{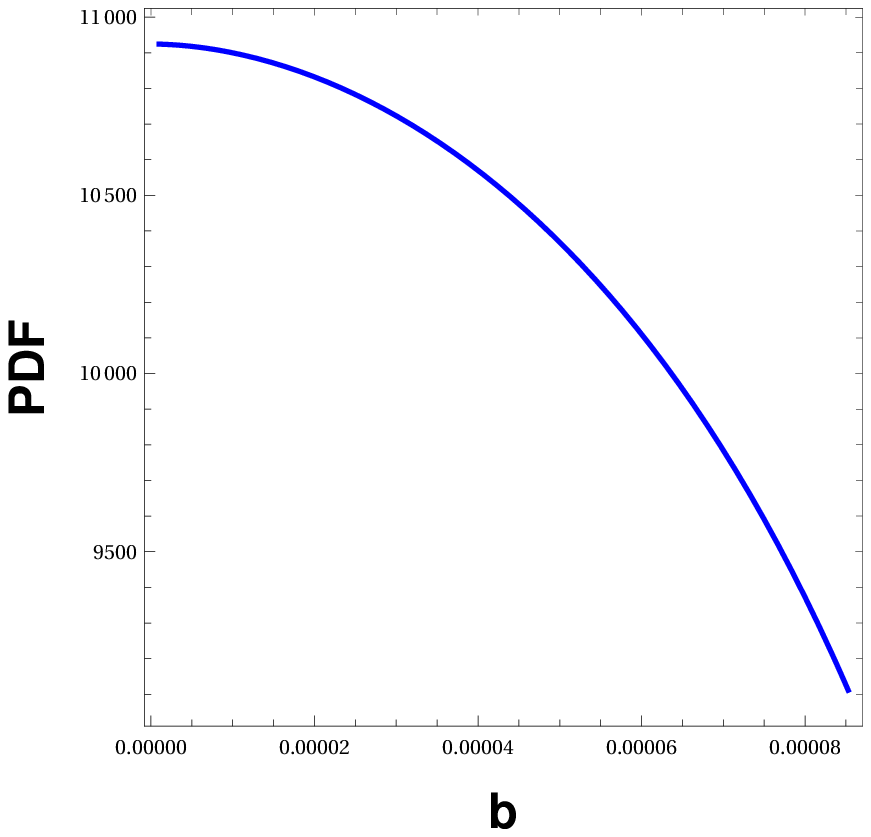}}
\begin{quotation}
\caption{\small The first CMB peak one-dimensional probability
distribution, after marginalizing on the other variables.}
\label{fig1}
\end{quotation}
\end{figure}

The position of the first peak in the CMB spectrum $l_1$ is related
to the acoustic scale $l_A$ by the relation
\beq
\n{1st-p}
l_1 = l_A (1 - \de_1),
\qquad \mbox{where} \qquad
\de_1 = 0.267 \left( \frac{\bar{r}}{0.3} \right)^{0.1},
\eeq
with $\,\bar{r} = \frac{\rho_r(z_{ls})}{\rho_m(z_{ls})}$ is evaluated
at the redshift of the last scattering surface, $z_{ls} = 1090$
\cite{Hu1994}. The acoustic scale is defined by
\beq
l_A \,=\, \frac{\pi \int_0^{z_{ls}} \frac{dz}{H(z)}}
{\int^\infty_{z_{ls}} \frac{c_s (z)}{c} \frac{dz}{H(z)}},
\eeq
where $c_s(z)$ is the sound speed
\beq
\n{so}
c_s (z) = c \left(
3 + \frac{9}{4} \frac{\Om^0_b}{\Om^0_\ga z}
\right)^{-1/2}.
\eeq
Here $\Om^0_b$ and $\Om^0_\ga$ stand for the present density
parameters of usual (baryonic) matter and photons, respectively.
The relation (\ref{1st-p}) does not depend on the dark energy model.
Here we consider the estimate $l_1 = 220.6 \pm 0.6$
and we use the values $\Om^0_\ga = 2.47 \times 10^{-5}/h^2$,
\ $\Om_b^0 = 0.022/h^2$ \ and \
$\Omega_r^0 = 4.18 \times 10^{-5}/{h^2}$ \ with the reduced
Hubble constant \ $h = 0.6732$ \cite{Planck2018}. Furthermore,
we let the free parameters run in the intervals
$\,\nu \in \big(0,\,10^{-4}\big)$, \
$b \in \big(0,\,10^{-4}\big)$ \ and
\ $\Om^0_{dm} \in \big( 0,\,0.95\big)$.
The minimization of the $\chi^2$ statistics is done according to
\beq
\n{chil1}
\chi^2_{l_1}
= \left[\frac{220.6 - l_1 (\Om^0_{dm},\nu,b)}{0.6} \right]^2,
\eeq
where this function has a local minimum around
\beq
\n{bf1}
 \Om^0_{dm} = 0.550,\qquad
\nu = 1.130\times 10^{-5},
\qquad
b = 4.117\times 10^{-5}.
\label{1-stdata}
\eeq

Here we can see that the current DM energy density value
$\Om_{dm}^0$ is higher than expected, indicating the necessity
of a more robust observational test to get a better fit with respect
to the standard model of cosmology (see Sec.~\ref{Data}).

It is easy to note that the value of $\Omega_{dm}^0$ quoted in
(\ref{1-stdata}) is dramatically different from the optimized value
in  $\Lambda$CDM. Certainly, this is not what should be expected
taking the relatively small values of DM warmness and running
into account. Indeed, the difference can be understood by the fact
that it corresponds only to the one particular observable, namely
the first CMB peak.
In this special situation, $b\neq 0$ (that indicates a WDM) implies
that more matter is required to reproduce the observed matter
agglomeration. In what follows, we will use a more complete
set of the observational data. Then the lower values for $b$ and
$\nu$ will be obtained, implying also a lower value for
$\Om_{dm}^0$, much closer to the conventional optimized value.
In particular, taking both $b$ and $\nu$ equal to zero, the usual
$\Lambda$CDM results are obtained.

\begin{figure}[t]
\vskip -0.5cm
\centerline{
\includegraphics[scale=0.65]{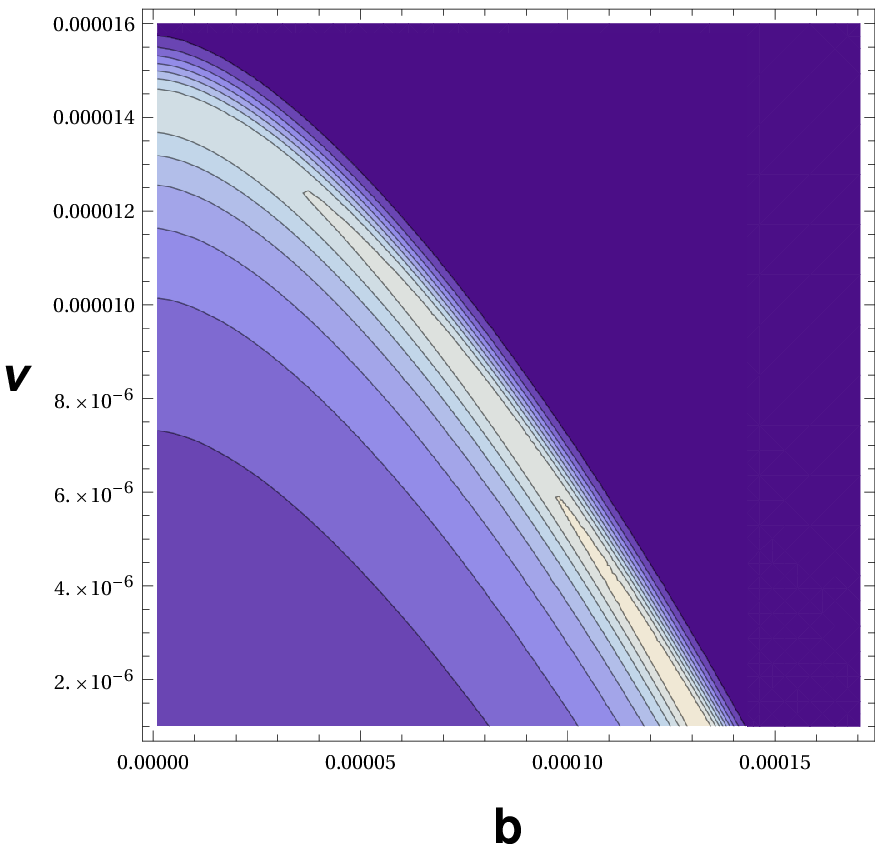}}

\centering{\includegraphics[scale=0.65]{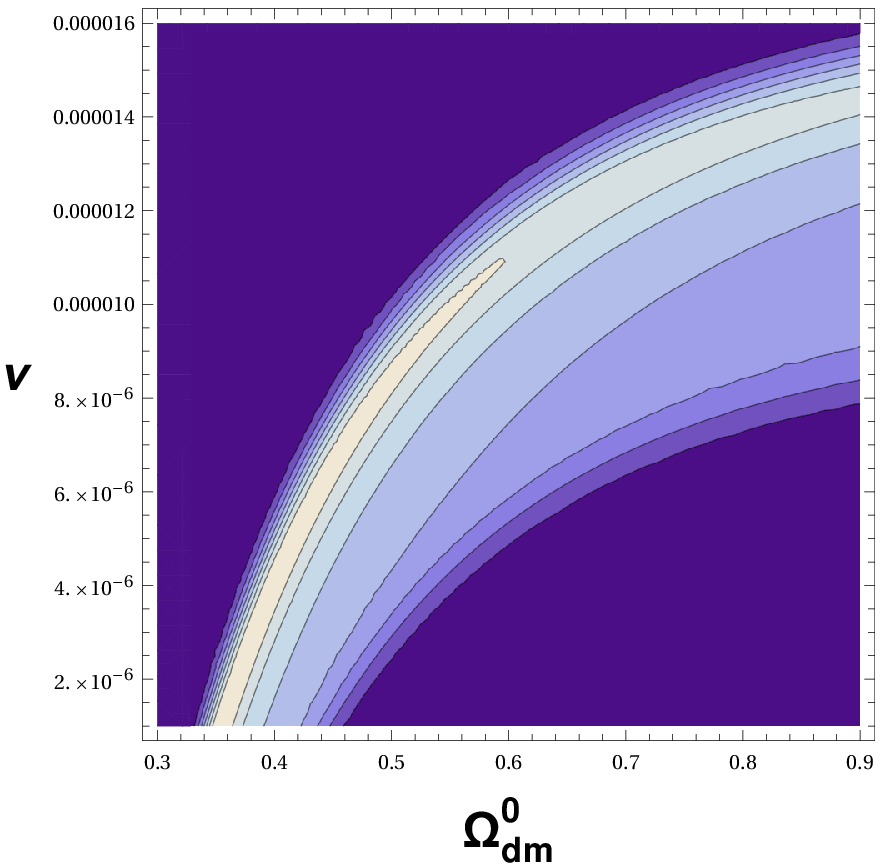}
\includegraphics[scale=0.65]{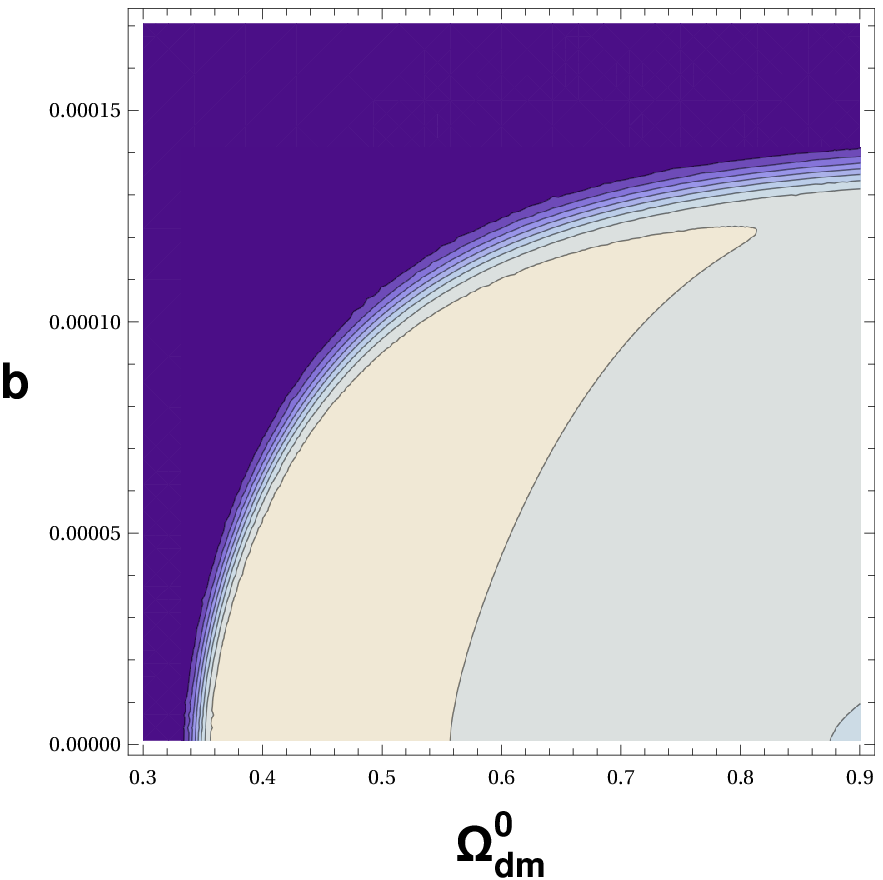}}
\begin{quotation}
\caption{\small
Two-dimensional probability distribution for the observational test using the first CMB acoustic peak. The brighter regions have higher
probabilities.}
\label{fig2}
\end{quotation}
\end{figure}

In Fig.~\ref{fig1} one can see the results
for the one-dimensional marginalized probability distribution
(PDF) for the free parameters of the model. It is easy to see
that this test alone cannot constraint too much the parameters.
Furthermore, the two-dimensional probability distribution,
with both parameters being varied and one is integrated
out, is shown in Fig.~\ref{fig2}. The regions of higher
probabilities in these plots are indicated by brighter tons.

The PDF distribution shown in this sub-section does not cover a compact and finite domain in the parameter space. This output of the numerical analysis is due to two reasons. First of all, it is due to the physical restriction on the parameters
of the model which we imposed. For example, we assumed that both
$\nu$ and $b$ should be positive and $\Omega_{dm}$ cannot be either
negative, neither greater than a threshold value. Certainly, from the
statistical point of view, this is odd, and hence we can not be
surprised by the unconventional form of the region in the parameter
space.

Second, it is known that for some specific models, a given parameter
may have a non-negligible PDF for disjoint regions. Even if such a feature may look unusual, it can be found in the literature. One particular example is the predictions for the equation of state
parameter $\alpha$ of the Generalised Chaplygin gas, where the
constraints from the Integrated Sachs-Wolfe (ISW) effect implies either
$\alpha \approx 0$ or $\alpha > 350$, with the limit
$\alpha \rightarrow \infty$ giving results similar to $\alpha = 0$
\cite{oliver}.

\subsection{Including BAO and SNIa data}\label{Data}

To find better constraints for our free parameters, in this section
it is constructed a more robust test using SNIa and BAO combined
data. Thus, we shall use
\beq
\chi^{2}_{total}=\chi^{2}_{l_1}+\chi^{2}_{BAO}+\chi^{2}_{SNIa},
\eeq
where $\chi^{2}_{BAO}$ and $\chi^{2}_{SNIa}$
are constructed following the reference \cite{Rivera2019}. The results
of this test are summarized in Table \ref{tab2} and are given by
\begin{equation}\label{bf2}
\bar{\Omega}_{m}^{0}=0.321,\qquad \bar\nu=2.442\times10^{-6},\qquad \bar{b}=1.888\times10^{-6}.
\end{equation}

Note that this $\Omega_{m}^0$ is lower than the previous estimate
(\ref{1-stdata}) and therefore looks closer to the constraints obtained
with the $\Lambda$CDM model.  It is also important to note
that in this combined test, it was used
$\Omega^0_m=\Omega^0_{dm}+\Omega^0_b$, instead of
$\Omega^0_{dm}$ as in section 2. Additionally, we observe
considerable variations in $\nu$ and $b$ values.

\begin{table}[t]
\centering
\begin{tabular}{@{}cccc@{}}
\toprule
Parameter        & SNIa & SNIa+BAO & SNIa+BAO+CMB\\
\midrule
$\chi_{min}^2$  & 562.227 & 583.289  & 585.440 \\[0.05cm]
$\Omega_{m}^0$ & 0.261 & 0.275  & 0.308 \\[0.15cm]
$\Omega_{\Lambda}^0$ & 0.680 & 0.647 & 0.656  \\[0.05cm]
\midrule
AIC  & 566.227 & 587.289 & 589.440  \\[0.05cm]
BIC  & 587.679 & 608.823 & 610.981  \\
\bottomrule
\end{tabular}
\caption{\small Summary of the observational constraints for the free
parameters and for the case of $\nu=b=0$ ($\Lambda$CDM with two
free parameters).}
\label{tab1}
\end{table}
\vspace{1cm}
\begin{table}[h]
\centering
\begin{tabular}{@{}cccc@{}}
\toprule
Parameter        & SNIa & SNIa+BAO  & SNIa+BAO+CMB\\
\midrule
$\chi_{min}^2$  & 562.315 & 583.402 & 585.374 \\[0.05cm]
$\Omega_{m}^0$  & 0.263 & 0.291 & 0.321 \\[0.05cm]
$\nu$  & 1.970 $\times 10^{-6}$ & 3.149$\times 10^{-6}$ & 2.442
$\times 10^{-6}$ \\[0.05cm]
$b$  & 1.170$\times 10^{-5}$ & 1.870$\times 10^{-5}$ & 1.888
$\times 10^{-6}$ \\[0.05cm]
\midrule
AIC         & 568.315 & 589.402 & 591.374 \\[0.05cm]
BIC  & 600.493 & 621.703 & 623.685  \\
\bottomrule
\end{tabular}
\caption{\small Summary of the observational constraints for the free
parameters and for the model RRG$+$RGE (with three free parameters:
$\Omega_{m}^{0}$, $\nu$ and $b$).}
\label{tab2}
\end{table}
\vspace{1cm}
\begin{table}[h!]
\centering
\begin{tabular}{@{}cccc@{}}
\toprule
Parameter        & SNIa & SNIa+BAO & SNIa+BAO+CMB\\
\midrule
$\Delta_{ik}$AIC        & 2.088 & 2.113 & 1,934 \\[0.05cm]
$\Delta_{ik}$BIC  & 12.814 & 12.880 & 12.704  \\
\bottomrule
\end{tabular}
\caption{\small Comparative analysis between the model RRG$+$RGE
(with three free parameters: $\Omega_{m}^{0}$, $\nu$ and $b$) and
the case of $\nu=b=0$ ($\Lambda$CDM with two free parameters).}
\label{tab3}
\end{table}
\newpage

\begin{figure} [t]
\vskip -0.5cm
\centerline{\includegraphics[scale=0.7]{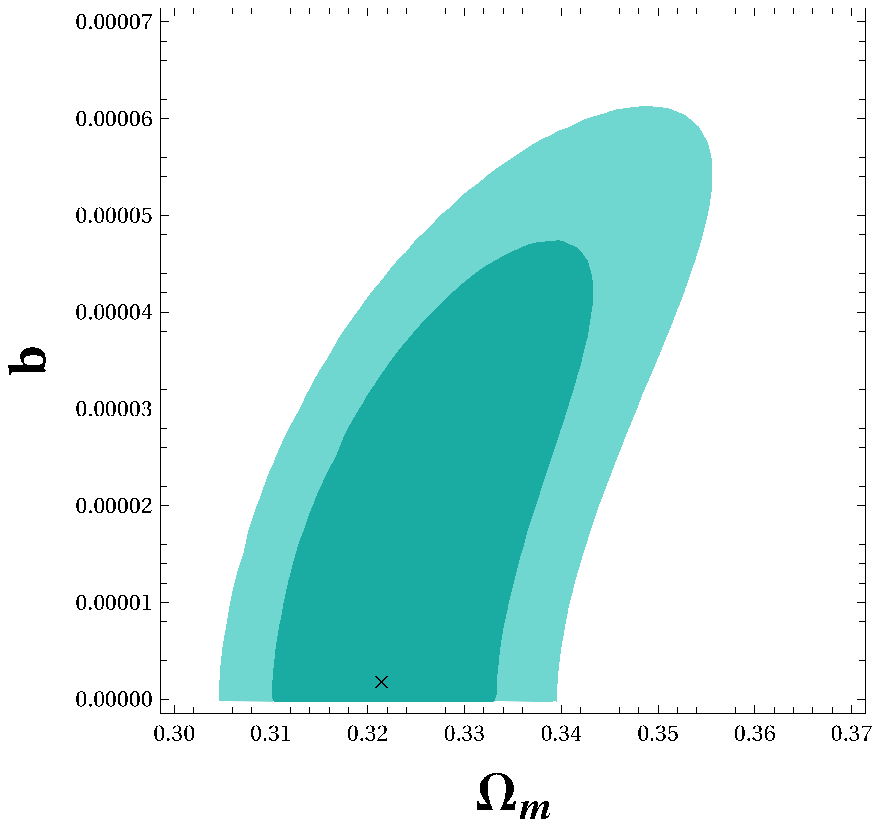}}
\centering{\includegraphics[scale=0.7]{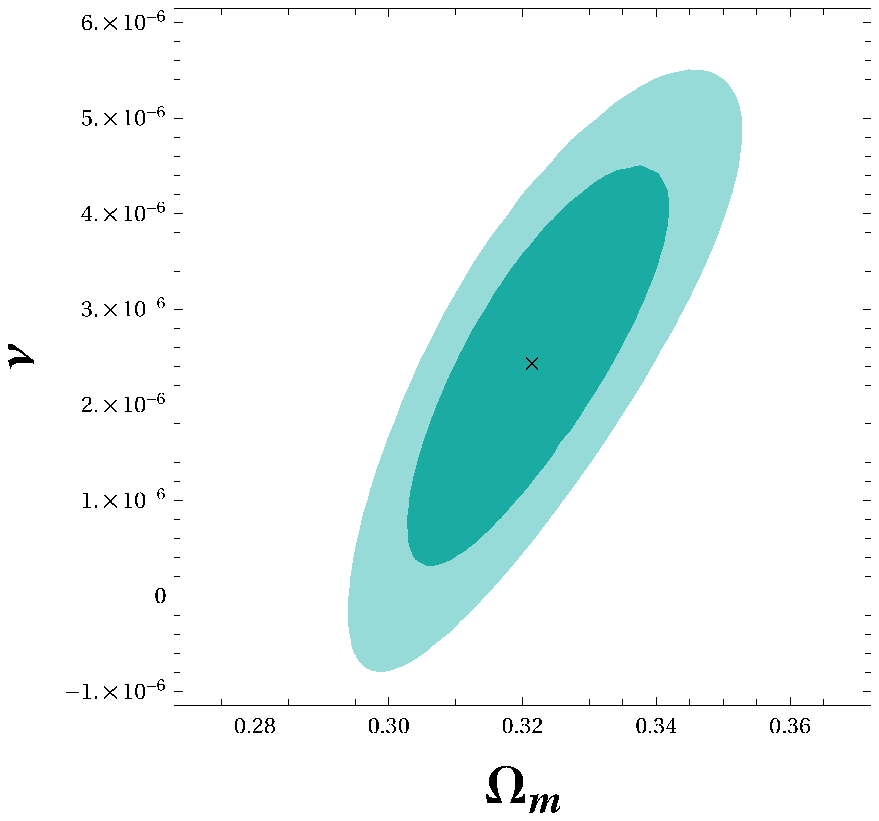}
\includegraphics[scale=0.7]{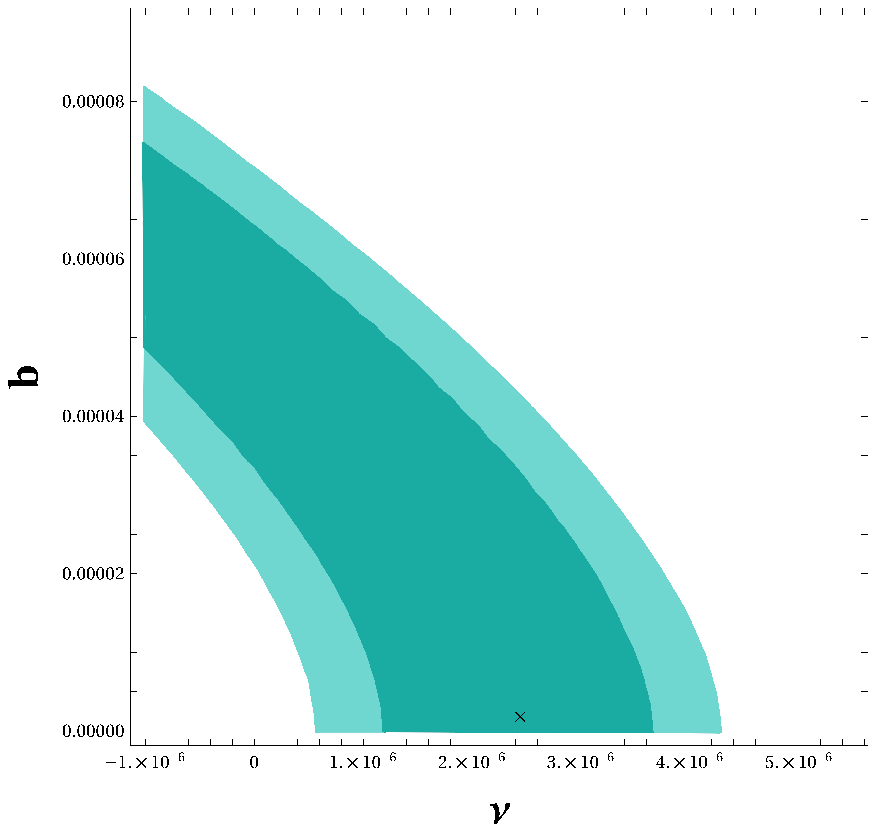}}
\begin{quotation}
\caption{\small Observational constraints for our three free
parameters $\nu$, $b$ and $\Omega_{m}^0$, for 1$\sigma$ and
2$\sigma$ levels. Here we have used SNIa+BAO+CMB combined
data. The marked points are given by
$(\bar{\Omega}_{m}^0,\bar{b})$, $(\bar{\Omega}_{m}^0,\nu)$
and $(\bar{\nu},\bar{b})$,
respectively, in correspondence with best fit values presented
above in Table \ref{tab2}.}
\label{fig5}
\end{quotation}
\end{figure}

On the other hand, in
Tables \ref{tab1} and \ref{tab3} we have written the best fit values for
$\Lambda$CDM ($\nu=b=0$) as our reference frame and the comparative
analysis between both models. Two of the main statistical criteria to
select models is the Akaike Information Criterion (AIC)
\cite{Akaike1974}, and the Bayesian Information Criterion (BIC)
\cite{Schwarz1978}, defined respectively by,
$AIC = \chi^2_{min} + 2\mu$ and $BIC = \chi^2_{min} + 2\mu\ln N$,
where $\mu$ is the number of degree of freedom and $N$ the number
of observational data. These criteria take into account the
number of free parameters of each model since the general tendency
of models of higher number of free parameters is to fit better the data.
Frequently it is used on the Jeffreys scale \cite{Liddle2007} to
quantify the relativity statistical relevance of the models. For the
AIC (BIC) parameter, models such $\Delta AIC (\Delta BIC) < 2$ have
strong support, weak support in the case $\Delta AIC (\Delta BIC) < 5$
and are disfavored for $\Delta AIC(\Delta BIC)  > 10$. From table
\ref{tab3} the Jeffreys scale applied to the AIC statistical criterion
favors the model RRG+RGE, while this model is strongly disfavored
using the BIC criterium, a consequence of the large number of
observational data, specially the SNIa data. This discrepancy in
using the two criteria is a common feature found in the literature
\cite{Szyd2008}-\cite{Szyd2015}. The contourplots with 1$\sigma$
and 2$\sigma$ levels for the CMB+BAO+SNIa combined data are
shown in Fig.~\ref{fig5}.

Let us note that the fact that the most probable values \eqref{bf1}
and \eqref{bf2}
include $\,\nu \neq 0\,$ does not constitute proof of the running of
the \CC.\ As usual, the statistics with an extra free parameter, such
as $\nu$, always gives the best values for the non-zero parameter,
and this is what we observe here. At the same time, it is remarkable
that letting \CC\ run does not lead to dramatic changes in the best
fit for other parameters, such as DM relative density $\Omega_{dm}^0$
and the warmness $b$.

\subsection{Matter power spectrum}

The matter power spectrum at $z = 0$ is given by
\beq
P(k) = | \de_m (k) |^2 = A k T^2(k)
\Big[\frac{\bar{g}(\Om_t^0)}{\bar{g}(\Om_m^0)} \Big]^2,
\eeq
where $A$ is a normalization constant of the spectrum. This constant
can be fixed from the spectrum of anisotropy of the CMB radiation and
\beq
\bar{g}(\Om) \,=\, \frac{5 \Om}
{2\,\big[\Om^{4/7} +1.01 (\Om/2+1) -0.75\big]}.
\eeq
Here we use the Bardeen-Bond-Kaiser-Szalay (BBKS) transfer
function \cite{BKS}
\beq
T(k) = \frac{\ln (1 + 2.34q)}{2.34q}
\big(1 + 3.89q + 16.1q^2 + 5.64q^3 + 6.71q^4\big)^{-1/4},
\eeq
where
\beq
q(k) = \frac{k}{h \Ga \mbox{Mpc}^{-1}}
\qquad \mbox{and} \qquad
\Ga \,=\, \Om^0_m h \,
\exp \Big\{-\Om^0_b - \frac{\Om^0_b}{\Om^0_m}\Big\},
\eeq
to construct a set of initial conditions for the system of equations
(\ref{1}), (\ref{2}), (\ref{3})-(\ref{5}).

\begin{figure}[t]
\centerline{\includegraphics[scale=1]{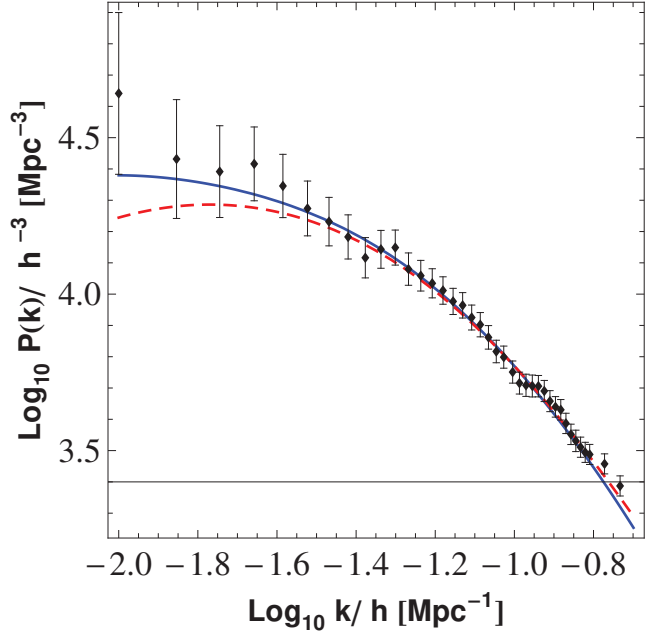}}
\begin{quotation}
\caption{\small Solid line: power spectrum of our model using
the best fit  of $\chi_{l_1}^2+\chi_{BAO}^2+\chi_{SNIa}^2$.
One can see that these values provide the linear power spectrum
which is compatible with the 2dFGRS data. Dashed line: power
spectrum obtained by BBKS transfer function with $\La$CDM
energy balance.}
\label{fig3.1}
\end{quotation}
\end{figure}

In Fig.~\ref{fig3.1} we compare the data from the 2dFGRS survey
\cite{2dFGRS} with the matter power spectrum of our model for the
energy balance obtained by the best fit  of
$\chi_{l_1}^2+\chi_{BAO}^2+\chi_{SNIa}^2$ (see Table
\ref{tab2}). Compared to more recent surveys (see e.g.
\cite{Parkinson:2012vd}), the 2dFGRG data present the
advantage of being less contaminated by
the standard model used in the calibration.

\section{Conclusions}
\label{s5}

We have implemented the model for the running of the cosmological
constant in an early stage of the Universe, where the dark matter
sector is modeled using the reduced relativistic gas model. At the
background level, the model was solved analytically, taking into
account the energy exchange between vacuum energy and usual
(baryonic) matter. The
effective equation of state parameter evolves as expected and we find
the best fit with respect to the standard model, once the constrained
values using SNIa+BAO+CMB combined data are obtained. Additionally,
this effective parameter goes to radiation value faster than in the
standard model for large redshift $z$. Besides the first CMB peak, the
SNIa and BAO data are used for constraining our free
parameters,obtaining a better correspondence with observations for this case.

On the other hand, when considering perturbations, to compute the
matter power spectrum, the system of equations for the geometric
perturbation, density contrasts, and velocities are found and solved
numerically.  We compared our results with the ones of the 2dFRG
data, obtaining a better correspondence for small $k$, in contrast to
the standard $\La$CDM model.

Our results suggest that a primordial running of the cosmological
constant and the possible creation of usual (baryonic) matter
particles at this early stage from vacuum energy, cannot be ruled
out and deserves more detailed exploration, e.g. in the possible
future work.

The model which we developed here explores the possibility that the
cosmological term decays into the baryonic component in the early
Universe, when the running of the cosmological constant and the
intensity of the gravitational field are sufficiently strong and, on the
other hand, baryonic matter contents can be regarded as
ultra-relativistic particles.
The parameter $\nu \neq 0$ indicates a non-constant cosmological
term and the parameter $b$ parameterizes the warmness of the matter
component.

The comparison with observation points to a small deviation from
the $\Lambda $CDM model as the preferred scenario, even though
the strict $\Lambda$CDM case, given by $\nu = 0$, is not excluded.
It must be remembered also that the running of the cosmological
term implies a new free parameter in comparison to the standard
cosmological model and, therefore, the results can not be interpreted
such that the statistical analysis proves that the cosmological
constant runs. Furthermore, the warmness of the dark matter
component $b \neq 0$ is allowed, with a present-day average speed
of the corresponding particles (or indefinite origin, as usual) of the
order of $10^{-5}\,c$.

Finally, let us stress the similarities and, on the other hand,
conceptual and technical differences between the model of running
cosmology which we dealt with in this work and the purely
phenomenological models describing the variable Dark Energy.
The model developed in this paper belongs to the class of
interaction models, where the energy-momentum tensor for
some components does not conserve separately as it happens
in the Standard Model. This means that a given component decays
into another one. This class of interacting model is nowadays very
popular in the study of the dark sector of the Universe, addressing
some questions like the coincidence problem. However, the
framework assumed here is quite different from most of these
papers. In the first place, we deal with an interacting model for
the early Universe, instead of a model for the late Universe. In
the present case, the (dynamical) cosmological term decays into
the usual (baryonic) matter when it is in the ultra-relativistic regime. On
the other hand, the form of the $H$-dependence for the cosmological
constant density in our model is defined from the quantum field
theory arguments \cite{CC-nova, DCCrun, PoImpo}. These arguments
defined the form of the IR running (\ref{CCrun}), leaving the
unique arbitrariness in the coefficient $\nu$.

From the technical side, it is interesting to see whether some known
phenomenological models describe an energy exchange between
vacuum and matter, like the one we considered here. Since there are
numerous models of this sort, the complete analysis is beyond our
possibilities, so we mention only one particular example. There is
some similarity with the model developed earlier in
Refs.~\cite{spindel1,spindel2} where it was considered the
energy exchange between vacuum and radiation, through the
evaporation of primordial black hole. In those references the form
of decaying of the cosmological term was fixed as an exponential
decay, leading to a smooth transition from inflation to a radiation
dominated phase, but with a prediction for the spectral index of
scalar perturbations was found to contradict the observational
constraints. This problem can be solved on the base of
Eq.~(\ref{CCrun}), by imposing upper bounds on the coefficient
$\nu$. On the other hand, the comparison with the scenario
described above is not direct since we consider a pos-inflationary
phase in contrast with the case treated in the mentioned references.

\section*{Acknowledgements}

\noindent
Authors are very grateful to Prof. A.A. Starobinsky for useful
critical observations.
J. A. Agudelo Ruiz thanks Prof. O.F. Piattella for useful discussions
during the lecture course on cosmology at UFES and to A. Bonilla
for his comments and discussion about data analysis and statistics.
He is also grateful to CAPES for supporting his Ph.D. project.
T.P.N. is grateful to CAPES for partial support and to the
N\'ucleo Cosmo-UFES and PPGCosmo for kind hospitality and
support during his visit at the Universidade Federal do Esp\'irito
Santo, where part of this work was done.
The work of J.F. was partially supported by Funda\c{c}\~{a}o
de Amparo \`{a} Pesquisa e Inova\c{c}\~{a}o do Esp\'{i}rito
Santo - FAPES and by the Conselho Nacional de
Desenvolvimento Cient\'{i}fico e Tecnol\'{o}gico - CNPq.
I.Sh. was partially supported by Conselho Nacional de
Desenvolvimento Cient\'{i}fico e Tecnol\'{o}gico - CNPq under the
grant 303635/2018-5 and Funda\c{c}\~{a}o de Amparo \`a Pesquisa
de Minas Gerais - FAPEMIG, under the project APQ-01205-16.



\end{document}